
\overfullrule=0pt

\input harvmac
\input epsf


\def\a{{\alpha}}
\def\l{{\lambda}}

\def\b{{\beta}}
\def\g{{\gamma}}
\def\d{{\delta}}

\def\N{{\nabla}}

\def\half{{1\over 2}}
\def\p{{\partial}}
\def\pb{{\bar\partial}}
\def\t{{\theta}}
\def\hat{\widehat}
\def\bar{\overline}

\def\nb {{\bar{\nabla}}}

\def\ah{{\widehat\alpha}}
\def\lh{{\widehat\lambda}}

\def\bh{{\widehat\beta}}

\def\th{{\widehat\theta}}
\def\L{{\Lambda}}
\def\CH{{\cal H}}

\def\ll{{\langle}}
\def\rr{{\rangle}}
\def\Tb{{\bf{T}}}
\def\lfa{{\leftrightarrow}}
\def\ln{{\rm ln}}
\def\irlog{{ln( {|k|^2\over \mu^2} )}}

\def\str{{{\rm Str}\,}}


\lref\berk{N. Berkovits,
JHEP 04 (2000) 018,
hep-th/0001035.}

\lref\adswitten{N. Berkovits, C. Vafa and E. Witten,
JHEP 9903 (1999)
018, hep-th/9902098.}

\lref\TseytlinMY{
  A.~A.~Tseytlin,
  Nucl.\ Phys.\  B {\bf 411}, 509 (1994)
  [arXiv:hep-th/9302083].
}

\lref\PulettiYM{
  V.~G.~M.~Puletti,
  JHEP {\bf 0809}, 070 (2008)
  [arXiv:0808.0282 [hep-th]].
}

\lref\MikhailovMR{
  A.~Mikhailov and S.~Schafer-Nameki,
  arXiv:0706.1525 [hep-th].
}

\lref\bersha{M. Bershadsky, S. Zhukov and A. Vaintrop, 
Nucl.Phys. B559 (1999) 205,
hep-th/9902180.}

\lref\bz{N. Berkovits, M. Bershadsky, T. Hauer, S. Zhukov and B. Zwiebach,
Nucl. Phys. B567 (2000) 61, hep-th/9907200.}

\lref\six{N. Berkovits, 
Nucl. Phys. B565 (2000) 333, hep-th/9908041.}

\lref\metsaev{R. Metsaev and A. Tseytlin, 
Nucl. Phys. B533 (1998) 109, hep-th/9805028.}

\lref\boer{
  J.~de Boer and K.~Skenderis,
  Nucl.\ Phys.\  B {\bf 481}, 129 (1996)
  [arXiv:hep-th/9608078].
}

\lref\chan{N. Berkovits and O. Chand\'{\i}a, 
Nucl. Phys. B596 (2001)
185, hep-th/0009168.}

\lref\sugra{N. Berkovits and P. Howe, 
Nucl.Phys. B635 (2002) 75, hep-th/0112160.}

\lref\wave{N. Berkovits, 
JHEP 0204 (2002) 037, hep-th/0203248.}

\lref\adscft{J. Maldacena, 
Adv.Theor.Math.Phys. 2 (1998) 231,
hep-th/9711200 ; S. S. Gubser, I. R. Klebanov and A. M. Polyakov,
Phys.Lett. B428 (1998) 105, hep-th/9802109 ; E. Witten,
Adv.Theor.Math.Phys. 2 (1998) 253,
hep-th/9802150.}

\lref\adsrev{O. Aharony, S. S. Gubser, J. Maldacena, H. Ooguri and Y. Oz,
Phys.Rept. 323 (2000) 183, hep-th/9905111.}

\lref\AdamWS{
  I.~Adam, A.~Dekel, L.~Mazzucato and Y.~Oz,
  JHEP {\bf 0706}, 085 (2007)
  [arXiv:hep-th/0702083].
}

\lref\loren{N. Berkovits and O. Chand\'{\i}a, 
Phys.Lett. B514 (2001) 394, hep-th/0105149.}

\lref\wadia{Gautam Mandal, Nemani V. Suryanarayana, Spenta R. Wadia,
Phys.Lett. B543 (2002) 81;
hep-th/0206103.}

\lref\oneloop{B. C. Vallilo, 
JHEP 0212 (2002) 042,
hep-th/0210064.}

\lref\abdalla{E. Abdalla, M. C. B. Abdalla and K. D. Rothe, 
World Scientific (Singapore) 2001.}

\lref\hidden{I. Bena, J. Polchinski, R. Roiban, 
hep-th/0305116.}

\lref\conf{
O.~Chandia and B.~C.~Vallilo,
JHEP {\bf 0404}, 041 (2004)
[arXiv:hep-th/0401226].
}

\lref\flatc{
B.~C.~Vallilo,
JHEP {\bf 0403}, 037 (2004)
[arXiv:hep-th/0307018].
}

\lref\oneloop{
B.~C.~Vallilo,
JHEP {\bf 0212}, 042 (2002)
[arXiv:hep-th/0210064].
}

\lref\fields{W. Siegel, {\it Fields,} hep-th/9912205.}

\lref\quantcon{  N.~Berkovits,
  JHEP {\bf 0503}, 041 (2005)
  [arXiv:hep-th/0411170].
}

\lref\twoloop{  B.~de Wit, M.~T.~Grisaru and P.~van Nieuwenhuizen,
  Nucl.\ Phys.\  B {\bf 408}, 299 (1993)
  [arXiv:hep-th/9307027].
}

\lref\LeeJY{
  K.~y.~Lee and W.~Siegel, 
  JHEP {\bf 0508}, 102 (2005)
  [arXiv:hep-th/0506198].
}

\lref\KnizhnikNR{
  V.~G.~Knizhnik and A.~B.~Zamolodchikov,
  Nucl.\ Phys.\  B {\bf 247}, 83 (1984).
}

\lref\BanksNR{
  T.~Banks and M.~B.~Green,
  JHEP {\bf 9805}, 002 (1998)
  [arXiv:hep-th/9804170].
}

\lref\KalloshQS{
  R.~Kallosh and A.~Rajaraman,
  Phys.\ Rev.\  D {\bf 58}, 125003 (1998)
  [arXiv:hep-th/9805041].
}

\lref\LeePA{
  K.~Lee and W.~Siegel, 
  arXiv:hep-th/0603218.
}

\lref\BergmanZH{
  O.~Bergman and S.~Hirano,
  arXiv:0902.1743 [hep-th].
}

\lref\BerensteinQD{
  D.~Berenstein and D.~Trancanelli,
  arXiv:0904.0444 [hep-th].
}

\lref\opes{ D. Nedel, O. Bedoya, D. Marchioro and B.C. Vallilo, {\it work in progress.}}

\lref\PolyakovQZ{
  A.~M.~Polyakov,
{  In *Les Houches 1988, Proceedings, Fields, strings and critical phenomena* 305-368. }
}

\lref\GomisJT{
  J.~Gomis, D.~Sorokin and L.~Wulff,
  JHEP {\bf 0903}, 015 (2009)
  [arXiv:0811.1566 [hep-th]].
}


\Title{ \vbox{\baselineskip12pt
\hbox{}}}
{{\vbox{\centerline{On the Non-renormalization of the AdS Radius}
\smallskip
\centerline{}
}}}
\centerline{Luca Mazzucato${}^{\heartsuit,\spadesuit}$
and Brenno Carlini Vallilo$^{\clubsuit,\spadesuit}$}
\bigskip
\centerline{\it $^{\heartsuit}$
Simons Center for Geometry and Physics}
\centerline{\it Stony Brook University, Stony Brook, NY 11794-3840, USA}
\centerline{\it $^{\clubsuit}$ Departamento de Ciencias F\'\i sicas,}
\centerline{\it Universidad Andres Bello, Republica 220, Santiago, Chile}
\centerline{$^\spadesuit$~~\it Kavli Institute for Theoretical Physics,}
\centerline{\it University of California at Santa Barbara, CA 93106-4030, USA}

\vskip .3in

\noindent We show that the relation between the 't Hooft coupling
and the radius of AdS is not renormalized at one-loop in the sigma model perturbation theory. We prove this by computing the quantum effective action for the superstring on $AdS_5\times S^5$ and showing that it does not receive any finite $\alpha'$ corrections. We also show that the central charge of the interacting worldsheet conformal field theory vanishes at one-loop.

\Date{June 2009}


\newsec{Introduction}

In this paper we consider string theory on an AdS background and discuss the quantum corrections to the target space radius, in the sigma model perturbation theory. The embedding of the string worldsheet into the target space is described by a sigma model on a (super)coset manifold. String propagation on the coset manifold $G/H$ can be described by a gauged WZW models. In bosonic WZW models, the level of the current algebra gets shifted at one-loop from $k$ to $k+\half c_G$, where $c_G$ is the quadratic Casimir of the group $G$ \KnizhnikNR. In the sigma model interpretation of WZW theory, the level is related to the radius of the target space
manifold, which is the inverse of the sigma model coupling constant. Therefore, the classical relation $R^2/\alpha'=k$ gets modified at one-loop to $R^2/\alpha'=k+\half c_G$ and in the full quantum theory there is a minimal value for the radius of the manifold, set by the quadratic Casimir of the group.
The situation is different for gauged WZW models with worldsheet supersymmetry \TseytlinMY. In that case, the fermionic and bosonic determinants cancel out and the relation between the radius and the level is not renormalized. These kinds of sigma model describe bosonic or RNS string theory on backgrounds supported by NS-NS flux. What happens  with Ramond-Ramond flux?

In this paper, we address this question in the case of superstring theory on $AdS_5\times S^5$, which is described by a sigma model on a supercoset. The AdS radius is again equal to the inverse of the sigma model coupling constant and is related to the 't Hooft coupling $\lambda$ of the dual ${\cal N}=4$ super Yang-Mills theory through the dictionary
$$
R^2/\alpha'=f(\lambda) \ , \qquad f(\lambda)\sim_{\lambda\to\infty}\sqrt{\lambda} + C_1+{\cal O}(1/\sqrt{\lambda})\ ,
$$
The leading term in the large 't Hooft coupling expansion corresponds to the classical supergravity dictionary, but in principle subleading terms are allowed and $C_1$ would arise at one-loop in the sigma model perturbation theory. This would be the analogue of the finite shift by $\half c_G$ in the level of the current algebra in gauged WZW models. In this paper, we will show that
$$
C_1=0 \ .
$$

The fact that the classical $AdS_5\times S^5$ solution of type IIB superstring is not modified by higher order $\alpha'$ corrections has been first discussed in the early days of AdS/CFT \BanksNR\KalloshQS. The first correction to type IIB supergravity comes at ${\cal O}(\alpha'^3)$ and it is the familiar $R^4$ term. The only component of the curvature that enters the $R^4$ term is proportional to the Weyl tensor, and since $AdS_5\times S^5$ is conformally flat, such leading correction vanishes. All the other terms related to $R^4$ by supersymmetry also vanish in this background, as well as the corresponding higher order corrections to the dilaton equation of motion. Using superspace techniques, due to the 32 supersymmetries of this background, this result can be extended to prove that the solution is not renormalized at all orders in $\alpha'$.\foot{The non-renormalization is also confirmed by explicit computations of the OPE's
of the currents \opes.} More recently, S-duality arguments applied to the giant magnon dispersion relation (where the function $f(\lambda)$ appears) have confirmed this result from the dual field theory side \BerensteinQD.

In order to study the renormalization of the radius, we need to compute the sigma model quantum effective action
$$
S_{eff}=S_{div}+S_{finite} \ .
$$
The divergent part of the effective action vanishes \oneloop\quantcon, which implies that the sigma model is conformally invariant and the radius does not run (see also \PulettiYM\MikhailovMR). However, one still needs to evaluate the finite part of the effective action, which may consist of local as well as non-local terms. The local terms can be reabsorbed or adjusted by local counter-terms to restore the classical symmetries. On the other hand, the presence of finite non-local contributions to the effective action could not be removed and would generate a non-zero $C_1$. Moreover, finite non-local terms in the effective action may produce gauge or BRST anomalies.

In this paper we will compute the finite part of the effective action at one-loop and show that all non-local contributions vanish.
Due to the presence of the  Ramond-Ramond flux, the worldsheet supersymmetric RNS description is not valid and we must use either the $\kappa$-symmetric Green-Schwarz-Metsaev-Tseytlin sigma model \metsaev\ or the BRST-symmetric pure spinor sigma model \berk. Since we would like to preserve covariance at all stages, we will consider the pure spinor approach. Because the covariant approach does not have worldsheet supersymmetry, we cannot borrow the RNS results, but we need to compute explicitly the one-loop effective action.

As a byproduct of our analysis, we will show that are no gauge nor BRST anomalies in the sigma model, confirming by an explicit one-loop computation the all-loop algebraic arguments in \quantcon. The last step in checking that the sigma model is quantum mechanically consistent at one-loop is the determination of its central charge, namely the leading quartic pole in the OPE of two stress tensors. Using the background field method, we show that the central charge vanishes.\foot{The same method used in this paper can be applied to the $AdS_p\times S^p$ pure spinor compactifications of \AdamWS, to obtain the same non-renormalization of the radius, due to the fact that the dual Coxeter numbers of the corresponding lower dimensional coset models vanish as well. For the non-critical $AdS_{2p}$ backgrounds in \AdamWS, the dual Coxeter number does not vanish and the radius may get renormalized already at one loop.}

Let us briefly comment on the case of the $AdS_4\times CP^3$ background. This is a solution of type IIA supergravity with only 24 supersymmetries, so the superspace arguments in \BanksNR\KalloshQS\ do not hold and one might expect the solution to get corrected. Even if the background is realized as the supercoset $Osp(4|6)/SO(1,3)\times U(3)$, where $Osp(4|6)$ has a vanishing dual Coxeter number just as $PSU(2,2|4)$, the full superstring sigma model is not described by a supercoset \GomisJT, unlike the $AdS_5\times S^5$ background. Hence, the methods we use in this paper may not be immediately generalized to that background. In \BergmanZH, a correction to the function $f(\lambda)=\sqrt{\lambda}(1+C_1/\sqrt{\lambda}+C_2/\lambda+\ldots)$ has been proposed, where $C_1=0$ and $C_2$ is a two loop numerical coefficient. It would be interesting to study this correction from the sigma model point of view.

In the rest of the introduction, we will review the computation of the effective action in the bosonic and RNS string. In section 2 and in the Appendix we collect some notations about the superstring sigma model on $AdS_5\times S^5$. In section 3 we compute the one-loop effective action using the background field method and discuss its properties. In section 4 we show that the central charge vanishes at one-loop.

\subsec{The Effective action in Bosonic and RNS String}

Let us review the computation of effective actions
in the closed bosonic and RNS string. We will set the notations and show why the bosonic string renormalizes, while worldsheet supersymmetry protects the metric from $\alpha'$ corrections at one-loop.

The bosonic string in a curved background is (we are
assuming that $B_{mn}=0$)
\eqn\boso{S_{bos}=\int d^2z \left[ \p x^m \pb x^n G(x)_{mn}\right].}
In the covariant background field expansion we fix a classical solution of the
worldsheet equations of motion $x_0$ and expand around
it in the quantum fluctuations $X$,
\eqn\bo{S_{bos}= S_0+\int d^2z  \eta_{ab}\left[\N X^a\nb X^b  + ... \right],}
where $...$ are terms depending on the curvature,
$\N X^a=\p X^a +  A^{ab}X_b$, $A^{ab}=\p x_0^m \omega_m^{ab}$ and
$\omega_m^{ab}$ is the spin connection. When one uses the normal coordinate expansion
within the background field method, local Lorentz invariance is used to
fix the spin connection to zero. In this case the resulting effective action
will not have this symmetry. Keeping local Lorentz invariance we have to
check if the effective action is not anomalous under this symmetry.

The effective action in momentum space is
\eqn\effbosomom{ S_{eff}= \half \int d^2k[ {1\over 2}A^{ab}(-k)A_{ab}(k)
{\bar k\over k} +
{1\over 2}\bar A^{ab}(-k)\bar A_{ab}(k){k\over \bar k} - {1\over 2}A^{ab}(-k)\bar A_{ab}(k)]}

The loop integrals are done using dimensional regularization adding a small
mass $m$ to the $X^a$ fields in order to regularize IR divergencies.
All the UV divergences cancel, and the dependence on the
dimensional regularization mass scale $\mu$ is an infrared effect, so
we can identify the mass regulator $m$ with $\mu$.\foot{We are ignoring IR divergent terms like $\irlog$. These terms are an IR effect and are expected to vanish when the
full perturbative series is summed \twoloop.}
The gauge variation of the effective action vanishes, even for the
non-local IR
divergent terms. We see that there is a finite local counter-term
responsible for the gauge invariance.
This is just a redefinition of the metric
$$G(x_0)_{mn}\to \widetilde G(x_0)_{mn}= G(x_0)_{mn} +
\alpha'{1\over 4}\omega_m^{ab}\omega_{n ab},$$
and the new metric now has a gauge transformation
$$\delta \widetilde G(x_0)_{mn} = \alpha'{1\over 4} \p_m \L_{ab}\omega_n^{ab}+
\alpha'{1\over 4} \p_n \L_{ab}\omega_m^{ab}.$$
The anomaly is trivial, this is the reason why we can fix the connection
to be zero when using normal coordinates.

Let us see what happens in RNS string. Its action in a curved
background is
\eqn\crns{S_{RNS}=\int d^2z\left[ (\p x^m \pb x^n +\half \psi^m\pb \psi^n +
\half \bar\psi^m\p\bar\psi^n + \right. }
$$\left. \half \psi^m \Gamma^n_{op}(x)\pb x^o\psi^p +
\half \bar\psi^m \Gamma^n_{op}(x)\p x^o\bar\psi^P)G(x)_{mn} +
{1\over 4}R_{mnop}\psi^m\psi^n\bar\psi^o\bar\psi^p \right].$$
Again, we fix a classical solution of the
worldsheet equations of motion $(x_0,\psi_0,\bar\psi_0)$ and expand around
it in the quantum fluctuations $(X,\Psi,\bar\Psi)$,
\eqn\rns{S_{RSN}= S_0+\int d^2z \left[\eta_{ab}\N X^a\nb X^b  +
\eta_{ab}\half\Psi^a \nb \Psi^b + \eta_{ab}\half\bar\Psi^a \N \bar\Psi^b
+...\right],}
where $...$ are terms depending on the curvature. The effective action is just
$$ -{1\over 2}\int d^2k[A^{ab}(-k)\bar A_{ab}(k)]$$
since non-local terms cancel due to worldsheet supersymmetry.
We have to add a local counter-term to cancel
the anomalous variation of this term, which removes the term above. Such term may appear naturally in other regularization scheme, see {\it e.g.} \PolyakovQZ.
We see that in the case of RNS superstring, even without gauge fixing
the connection, there are no finite local corrections in the effective action. In RNS the IR divergent terms are also present and are gauge
invariant. These terms are related to the fact that $X(z)$ is not a primary
field in two dimensions.
In the case case of Type I or Heterotic string we would have the usual
local Lorentz anomaly that appears because we have only left moving fermions,
which can be canceled by a variation of the $B$ field.
In coset models, like the superstring in $AdS_5\times S^5$ space, it is
useful to keep the connection unfixed since this
simplifies significantly the background field expansion.

\newsec{Pure Spinor Superstring in the $AdS_5\times S^5$ Background}

The $AdS_5\times S^5$ background can be described by the coset superspace
$PSU(2,2|4)/SO(4,1)\times SO(5)$ \metsaev.
From the metric and structure constants listed in Appendix A,
we see that the super Lie algebra $PSU(2,2|4)$ admits a decomposition \bz\ under ${\bf Z_4}$,
$\CH=\sum \CH_i$, $i=0$ to $3$ \eqn\dec{\Tb_{\underline a} \in
\CH_2, \quad \Tb_{[\underline{ab}]} \in \CH_0, \quad \Tb_\a \in \CH_1,
\quad \Tb_\ah \in \CH_3.}

Using the supertrace notation and setting $\alpha'=1$, we write
the $AdS_5\times S^5$ pure spinor action \chan\oneloop\ as
\eqn\class{S_0 = {R^2\over 2\pi}\int d^2z\,\str\left(\half J_2 \bar J_2 + {3\over 4}
J_3 \bar J_1  + {1\over 4}
J_1 \bar J_3
+
w  \bar\nabla \lambda  +
\widehat w \nabla \lh  - N \widehat N \right)}
where
\eqn\notat{J_0 = (g^{-1}\p g)^{[ab]} \Tb_{[ab]},\quad
J_1 = (g^{-1}\p g)^{\a} \Tb_{\a},\quad
J_2 = (g^{-1}\p g)^{m} \Tb_{m},\quad
J_3 = (g^{-1}\p g)^{\ah} \Tb_{\ah},}
$$
w = w_\a \Tb_\ah \d^{\a\ah},
\quad\l = \l^\a \Tb_\a,\quad N =  -\{w,\l\}, $$
$$\overline J_0 = (g^{-1}\overline \p g)^{[ab]} \Tb_{[ab]},\quad
\overline J_1 = (g^{-1}\overline \p g)^{\a} \Tb_{\a},\quad
\overline J_2 = (g^{-1}\overline \p g)^{m} \Tb_{m},
\quad \overline J_3 = (g^{-1}\overline \p g)^{\ah} \Tb_{\ah}, $$
$$\widehat w = \widehat w_\ah \Tb_\a \d^{\a\ah},
\quad\lh = \lh^\ah \Tb_\ah,\quad \widehat N =
-\{\widehat w,\lh\},$$
$$\nabla  = \partial  + [J_0, \;\; ], \quad
\bar\nabla  = \bar\partial  + [\bar J_0,\;\; ], $$
$ \d_{\a\bh} = (\g^{01234})_{\a\bh}$,
$\Tb_A$ are the $PSU(2,2|4)$ Lie algebra generators. Note that
\eqn\metric
{\{\Tb_\a,\Tb_\b\} = \g_{\a\b}^m \Tb_m,\quad
\{\Tb_\ah,\Tb_\bh\} = \g_{\ah\bh}^m \Tb_m,\quad
\{\Tb_\a,\Tb_\bh\} = (\half \g^{[ab]}\g^{01234})_{\a\bh} \Tb_{[ab]} \ .}
Also note that $\l$ and $\lh$ are fermionic since $(\Tb_\a,\Tb_\ah)$
are fermionic
and $(\l^\a,\lh^\ah)$ are bosonic.
The action of \class\ is manifestly invariant under global $PSU(2,2|4)$
transformations which transform $g(x,\t,\th)$ by left multiplication
as $\d g = (\Sigma^A T_A) g$ and is also manifestly invariant
under local $SO(4,1)\times SO(5)$ gauge transformations which transform
$g(x,\t,\th)$ by right multiplication as $\d_\Lambda
g= g\Lambda$ and transform
the pure spinors as
$$\d_\L \l = [\l,\L],\quad \d_\L\lh = [\lh,\L],\quad
\d_\L w = [w,\L],\quad \d_\L\widehat w = [\widehat w,\L]$$
where $\L = \L^{[ab]} T_{[ab]}$.

\newsec{Effective Action}

We can quantize the classical action \class\ using the covariant
background field method. This method was used in \oneloop\ to prove
one-loop conformal invariance of \class. In this section we will
compute the one-loop effective action.

\subsec{Matter}

A classical background field
$\tilde g$ is
chosen and the quantum fluctuations are parameterized by
$X=X_1+X_2+X_3$, with $g=\tilde g e^{X}$, in the gauge $X_0=0$.
The quantum currents are \eqn\currents{ J = g^{-1}\p g =
e^{- X}\tilde J e^{ X}+e^{- X}\p e^{ X},}
$$ \bar J = g^{-1}\bar \p g = e^{- X}\tilde{\bar J} e^{ X}+
e^{- X}\bar\p e^{ X},$$
where $\tilde J=\tilde g^{-1}\p \tilde g$.

The OPE for the quantum fluctuations is
\eqn\opesx{X^A(z)X^B(w)\to  -\eta^{BA}\ln|z-w|^2 \ .
}
Since we are going to do a one-loop computation, we  expand
\currents\ up to the second order in $X$,
\eqn\expansion{\eqalign{
J|_i=&\tilde J|_i+{1\over R}\left(dX+[\tilde J,X]\right)|_i+{1\over 2R^2}\left[dX+[\tilde J,X],X\right]|_i +{\cal O}(R^{-3}) \ ,
 }}
We separate the relevant terms in the action as follows. The kinetic terms are
\eqn\kinaction{
S_{\rm kin}=\int d^2z\str\left( \half \nabla X_2\bar \nabla X_2+{1\over4}\nabla X_1\bar \nabla X_3+{3\over 4}\nabla X_3\bar \nabla X_1\right) \ ,
}
where the covariant derivative $\nabla X_i=\partial X_i+[J_0,X_i]$ depends on the background gauge current.
We will put in $S_I$ all the terms that contain either $J_2$ or $\bar J_2$ or both
\eqn\sone{\eqalign{
S_I=&\int d^2z\str\Bigr( \half J_2[X_1,\bar\nabla X_1]+\half \bar J_2[X_3,\nabla X_3]\cr&
+{1\over4}J_2\left[[\bar J_2,X_1],X_3\right]-{1\over4}J_2\left[[\bar J_2,X_3],X_1\right] \Bigr)\ .
}}
To ease the notation, we will drop the $\tilde{}$ on top of the background currents.
We will put into $S_{II}$ all the terms that depend on $J_1$ or $\bar J_3$ or both
\eqn\stwo{\eqalign{
S_{II}=&\int d^2z\str\Bigr( {1\over 8}J_1(3[X_1,\bar \nabla X_2]+{5}[X_2,\bar \nabla X_1])+{1\over 8}\bar J_3(3[X_3, \nabla X_2]+{5}[X_2,\nabla X_3])\cr
& -\half J_1\left[[\bar J_3,X_2],X_2\right]+{1\over4}J_1\left[[\bar J_3,X_1],X_3\right]-{1\over 4}J_1\left[[\bar J_3,X_3],X_1\right] \Bigr) \ .
}}
Finally, we collect in $S_{III}$ the terms that depend on $J_3$ or $\bar J_1$ or both
\eqn\sthree{\eqalign{
S_{III}=&\int d^2z\str\Bigl( {1\over8}\bar J_1([X_1,\nabla X_2]-[X_2,\nabla X_1])+{1\over 8}J_3([X_3,\bar\nabla X_2]-[X_2,\bar\nabla X_3])\cr
& +\half\bar J_1\left[[J_3,X_2],X_2\right]+{3\over4}\bar J_1\left[[ J_3,X_1],X_3\right]+{1\over 4}\bar J_1\left[[J_3,X_3],X_1\right] \Bigr) \ .
}}

\subsec{Ghost}

Let us consider the ghost part of the one loop effective action. We expand the left and right moving ghosts into upper case background fields and lower case fluctuations
\eqn\ghostexp{
(w,\lambda)\to(W+w,L+\lambda)\ ,\qquad (\hat w,\hat \lambda)\to(\hat W+\hat w,\hat L+\hat \lambda)\ .
}
The ghost Lorentz currents are expanded as
\eqn\lorentz{
N\to N_{(0)}+{1\over R}N_{(1)}+{1\over R^2}N_{(2)} ,\qquad
\hat N\to \hat N_{(0)}+{1\over R}\hat N_{(1)}+{1\over R^2}\hat N_{(2)} \ ,
}
where $(N_{(0)},\hat N_{(0)})$ denote the background currents while
\eqn\fluctuan{\eqalign{
N_{(1)}=-\{W,\l\}-\{w,L\}\ ,\qquad &N_{(2)}=-\{w,\l\}\ ,\cr
\hat N_{(1)}=-\{\hat W,\hat \l\}-\{\hat w,\hat L\}\ ,\qquad & N_{(2)}=-\{\hat w,\hat\l\} \ .
}}
We expand the classical ghost action  according to \fluctuan\ and collect the terms quadratic in the fluctuations
\eqn\ghostactio{\eqalign{
S=&\half\int d^2z\str\,\Bigl\{
N_{(0)}\left([\bar\nabla X_3,X_1]+[\bar \nabla X_2,X_2]+[\bar\nabla X_1,X_3]\right)\cr&+\hat N_{(0)}\left([\nabla X_3,X_1]+[\nabla X_2,X_2]+[\nabla X_1,X_3]\right)\cr&-N_{(1)} \hat N_{(1)}+N_{(2)}(\bar J_0-\hat N_{(0)})+(J_0-N_{(0)})\hat N_{(2)}\Bigr\} \ .
}}

\subsec{Effective action}

The computation of the effective action at one loop order proceeds as follows. There are two kind of terms that we need to compute, schematically
\eqn\effact{
S_{eff}=\int d^2z\langle {\cal L}(z)\rangle-\half \int d^2z\int d^2w\langle {\cal L}(z){\cal L}(w)\rangle \ ,
}
where the $\langle\cdot\rangle$ denotes functional integration over the fluctuating fields. The first term $\langle {\cal L}(z)\rangle$ corresponds to the normal ordering of the composite operators in the lagrangian: it is just given by the one loop self energy of the fluctuations at the same point, in operators with two external currents. These are the second lines in \sone, \stwo\ and \sthree. The second term $\langle {\cal L}(z){\cal L}(w)\rangle $ corresponds to the one loop fish diagram generated by the contraction of the operators with one external current, namely the first lines in \sone, \stwo\ and \sthree.

As an example, we explicitly evaluate the term in the effective action proportional to the operator $\str J_1\bar J_3$. Using the OPE's for the quantum fluctuations \opesx\ we find
 \eqn\leffII{\eqalign{
{S}_{eff}=&\int d^2z\,J_1^\a \bar J_3^\ah (z)[-\ln(0)]\left({3\over4}f_{\ah a}^\b f_{\a\b}^a+{1\over4}f_{\ah \b}^{[ef]}f_{\a[ef]}^\b\right)\cr
&+\int d^2z\int d^2w\,J_1^\a(z)\bar J_3^\ah(w)f_{\ah a}^\b f_{\a\b}^a\left({34\over 64}\delta^{(2)}(z-w)\ln|z-w|^2-{30\over 64}{1\over|z-w|^2}\right) \ ,
}}
The last line in \stwo\ contributes to the self-energy graphs
in the first line, while the first line in \stwo\ contributes through the OPE of ${3\over8}[X_1,\bar\nabla X_2]+{5\over8}[X_2,\bar\nabla X_1]$ and ${3\over8}[X_3,\nabla X_2]+{5\over8}[X_2,\nabla X_3]$, generating the term in the second line. Using the map to momentum space, listed in Appendix B, this gives the following term in the effective action
\eqn\selfe{
S^{(1)}_{eff}=\int d^2z\ J_1^\alpha \bar J_3^{\ah}\left[ 1\over{\epsilon}\right]\left({3\over4}f^\beta_{\ah a}f^a_{\a\b}+{1\over4}f^{[ef]}_{\ah\b}f^\b_{\a[ef]}\right) -\int d^2z J_1^\alpha\bar J_3^{\ah}f_{\ah a}^\b f_{\a\b}^a\left[1 +  {1\over{\epsilon}} \right] \ .
}
The divergent part of the effective action cancels and we are left with the following finite piece\foot{We used the identity $f^\b_{\ah a}f^\ah_{\a\b}=-f_{\ah\bh}^a f_{\a a}^\bh$ and the fact that the combination $\half f_{\ah a}^\b f_{\a\b}^a-\half f_{\ah \b}^{[ef]}f^\b_{\a[ef]}=R_{\ah\a}(G)=0$, since $G=PSU(2,2|4)$ has vanishing dual Coxeter number. Moreover we used the identity $\gamma^m_{\ah\bh}=\eta_{\alpha\ah}\eta_{\beta\bh}(\gamma^m)^{\alpha\beta}$.} \eqn\finiteac{
S_{eff}(J_1\bar J_3)= 10\int d^2z \str J_1\bar J_3\ ,
}
which is local. By analogous computations we end up with the full effective action
\eqn\effective{
\eqalign{
S_{eff}=\int d^2z\,\str \Bigl(&a_1J_2\bar J_2+a_2 J_1\bar J_3+\half c_2(H)(J_0\bar J_0-N\bar J_0-\hat N J_0) \Bigr)\ .
}}
where $a_1=8$, $a_2=10$ and $c_2(H)=3$ is the quadratic Casimir of the gauge group $H=SO(1,4)\times SO(5)$. We did not include the IR singular terms proportional to $\ln |p|^2/\mu^2$, which are expected to vanish once the full perturbative series is included \twoloop. The expression \effective\ has the following properties:
\item{i)} It is local, hence it can be removed by adding a local counter-term $S_c=-S_{eff}$ to the action, according to the prescription given in \quantcon\ to preserve gauge and BRST symmetries. As a result, we proved that there are no gauge nor BRST anomalies at one-loop.
\item{ii)} By explicit computation, we checked that operators of weight $(2,0)$ and $(0,2)$, {\it e.g.} of the kind $\str J_i J_j$, $\str NN$ or $\str \hat N\hat N$, are not generated at one-loop, due to remarkable cancellations in the diagrams caused by the vanishing of the dual Coxeter number of $PSU(2,2|4)$.

\newsec{Central Charge}

The last step in checking that the worldsheet theory is consistent at one-loop is the computation of the central charge. The stress tensor for the action \class\ is \eqn\stresste{
T=-{\rm Str}\left( \half J_2\bar J_2+J_1J_3+w\nabla \lambda \right)\ .
}
We want to compute the one loop correction to the central charge. This is the quartic pole in the OPE
$$
\langle T(z) T(0) \rangle= {c/2 \over z^4}+\ldots ,
$$
where $\langle\cdot\rangle$ denotes functional integration.
We expand $T$ according to \currents\ and we compute the contractions of the fluctuations. The terms coming from the action do not contribute to the central charge. We find a leading tree level contribution, proportional to $1/R^4$, where $R$ is the radius and the action is normalized as $S={R^2\over 2\pi}\int {\cal L}$. The one loop correction is proportional to $1/R^6$. To compute terms of order $1/R^8$ we need to expand \currents\ up to ${\cal O}(R^{-3})$, so they will be neglected and we will stop at one loop. We find
\eqn\twotwo{
\langle \half {\rm Str} J_2J_2(z)\half {\rm Str} J_2J_2(0)\rangle={1\over R^4}{1\over z^4}\left({10\over2}-{1\over 2R^2} [1+\ln|z-w|^2]\eta^{lm} f_{l\alpha}^{\hat \delta}f_{m\hat\delta}^\alpha\right) \ ,
}
where $\eta^{lm} f_{l\alpha}^{\hat \delta}f_{m\hat\delta}^\alpha={1\over4}{\rm Tr}\gamma^a\gamma_a=40$.
The first term arises from the double contraction at tree level, while the second comes from the triple contraction at one loop.
The second contribution is
\eqn\onethree{
\langle {\rm Str} J_1J_3(z){\rm Str} J_1J_3(0)\rangle=-{1\over R^4}{1\over z^4}\left({32\over2}+{1\over 2R^2} [1+\ln|z-w|^2]\eta^{lm} f_{l\alpha}^{\hat \delta}f_{m\hat\delta}^\alpha\right) \ .
}
The mixed term is
\eqn\mixed{
-\langle {\rm Str}  J_2J_2(z){\rm Str} J_1J_3(0)\rangle={1\over R^6}{1\over z^4} [1+\ln|z-w|^2]\eta^{lm} f_{l\alpha}^{\hat \delta}f_{m\hat\delta}^\alpha\ .
}
By summing up \twotwo, \onethree\ and \mixed\ we get the total contribution of the matter part. The one-loop correction cancels out exactly, leaving only the tree level part, which is the same as in flat space
\eqn\matter{
\langle T_{matter}(z) T_{matter}(0)\rangle = -{1\over R^4}{22\over 2z^4} \ .
}

Let us look at the ghost part. The tree level contribution involves a trace on the ghost spinor indices and is equal to the analogous flat space contraction. In the gauge $X_0=0$ the ghost sector does not give any one-loop correction and it starts contributing only at two loops (the leading term in $w [J_0,\lambda]$ with no external fields is ${\cal O}(R^{-4})$), so we find
\eqn\ghostt{
\langle T_{gh}(z)T_{gh}(0)\rangle ={1\over R^4}{22\over 2z^4}\ ,
}
and by adding \ghostt\ and \matter\ we proved that the total central charge vanishes at one loop.

Since the effective action does not receive any finite corrections at one loop, there is no correction to the stress tensor either.

\vskip 15pt
{\bf Acknowledgements:} We would like to thank Daniel Nedel, Yaron Oz, Leonardo Rastelli and Matthias Staudacher for valuable
discussions. We also thank Nathan Berkovits for collaboration during the initial stages of this work. The authors would like to thank the KITP at Santa Barbara for the kind hospitality, where this project has started. This research was supported in part by DARPA under Grant No. HR0011-09-1-0015 and by the National Science Foundation under Grant No. PHY05-51164.

\appendix{A}{Notations}

The metric in the $PSU(2,2|4)$ is

\eqn\metric{\eta^{\underline{ab}}=\eta^{\underline{ba}};\quad
\eta_{\underline{ab}}=\eta_{\underline{ba}};}
$$\eta^{\a\bh}=-\eta^{\bh\a}=(\gamma_{01234})^{\a\bh};\quad
\eta_{\a\bh}=-\eta_{\bh\a}=(\gamma_{01234})_{\a\bh};$$
$$\eta^{[ab][cd]}={1\over 2}\eta^{a[c}\eta^{d]b};\quad
\eta^{[a'b'][c'd']}=-{1\over 2}\eta^{a'[c'}\eta^{d']b'};$$
$$\eta_{[ab][cd]}={1\over 2}\eta_{a[c}\eta_{d]b};\quad
\eta_{[a'b'][c'd']}=-{1\over 2}\eta_{a'[c'}\eta_{d']b'}.$$
The underlined vector index $\underline{a}=0,\ldots,9$ is ten-dimensional, while $a=0,\ldots,4$ and $a'=5,\ldots,9$ represent the $AdS_5$ and the $S^5$ directions respectively; the indices $\alpha,\hat \alpha=1,\ldots,16$ describe  $SO(1,9)$ Weyl spinors of the same chirality. Capital letters are collective $PSU(2,2|4)$ indices, $A=(\underline{a},\alpha,\hat\alpha,[\underline{a}\underline{b}])$.

The metric satisfies $\eta^{AB}\eta_{BC}=\delta^A_C$. Denoting
$(\Tb_{\underline{a}}, \Tb_{[\underline{ab}]}, \Tb_{\a}, \Tb_{\ah})$ the
generators of the algebra, $\eta_{AB}= \ll \Tb_A , \Tb_B\rr$.
The non-vanishing structure constants $f_{AB}^C$
of the $PSU(2,2|4)$ algebra are
\eqn\structure{f_{\a\b}^{\underline{c}} =\g^{\underline{c}}_{\a\b},\quad
f_{\ah\bh}^{\underline{c}} =\g^{\underline{c}}_{\ah\bh},
}
$$
f_{\a \bh}^{[ef]}=
\half(\g^{ef})_\a{}^\g \eta_{\g\bh},\quad
f_{\a \bh}^{[e'f']}=-\half
(\g^{e'f'})_\a{}^\g \eta_{\g\bh},$$
$$f_{\a \underline{c}}^\bh
=-(\g_{\underline c})_{\a\b}
\eta^{\b\bh},\quad
f_{\ah \underline{c}}^\b =
(\g_{\underline c})_{\ah\bh} \eta^{\b\bh},$$
$$f_{c d}^{[ef]}= \d_c^{[e} \d_d^{f]},
\quad f_{c' d'}^{[e'f']}= -\d_{c'}^{[e'} \d_{d'}^{f']},$$
$$f_{[\underline{cd}][\underline{ef}]}^{[\underline{gh}]}=\half (
\eta_{\underline{ce}}\d_{\underline{d}}^{[\underline{g}}
\d_{\underline{f}}^{\underline{h}]}
-\eta_{\underline{cf}}\d_{\underline{d}}^{[\underline{g}}
\d_{\underline{e}}^{\underline{h}]}
+\eta_{\underline{df}}\d_{\underline{c}}^{[\underline{g}}
\d_{\underline{e}}^{\underline{h}]}
-\eta_{\underline{de}}\d_{\underline{c}}^{[\underline{g}}
\d_{\underline{f}}^{\underline{h}]})$$
$$f_{[\underline{cd}] \underline{e}}^{\underline{f}} = \eta_{\underline{e}
\underline{[c}} \d_{\underline d]}^{\underline{f}},\quad
f_{[\underline{cd}] \a}^{\b} = \half(\g_{\underline{cd}})_\a{}^\b,\quad
f_{[\underline{cd}] \ah}^{\bh} = \half(\g_{\underline{cd}})_\ah{}^\bh.$$

\appendix{B}{Map to momentum space}

It is not clear upon inspection which terms in the effective action in coordinate space are finite or divergent. Also, we have
the usual complications due to infrared divergencies. To clarify the interpretation, we will transform
the above two point functions into loop integrals in the momentum space. We
perform the loop integral using dimensional regularization adding a small
mass $m$ to the $X^a$ fields in order to regularize IR divergencies. The dependence on the
dimensional regularization mass scale $\mu$ is an infrared effect, so
we can identify the mass regulator $m$ with $\mu$.
In order to
simplify the calculation, we build a dictionary between the above two point
functions and and the corresponding result of the integration over the
momenta \boer\
\eqn\dictio{ \eqalign{{1\over{(z-w)^2}} &\quad\leftrightarrow \quad -{ p \over{\bar p} }\ ,\cr
{1\over{(\bar z-\bar w)^2}} &\quad\leftrightarrow\quad -{\bar p \over p} \ ,\cr
{\ln|z-w|^2\over{(z-w)^2}} &\quad\lfa \quad -{ \bar p \over p}(1+
2\ln( {|p|^2\over \mu^2} ))\ ,\cr
{\ln|z-w|^2\over{(\bar z-\bar w)^2}} &\quad\lfa \quad -
{ p \over{\bar p}}(1 + 2\ln( {|p|^2\over \mu^2} ))\ ,\cr
{1\over|z-w|^2}&\quad\lfa\quad 1+{1\over \epsilon} -2\ln( {|p|^2\over \mu^2} ) \ ,\cr
-\ln|z-w|^2\delta(z-w) &\quad\lfa\quad 1+{1\over \epsilon} \ ,\cr
-\ln(0) &\quad\lfa\quad {1\over \epsilon}\ .
}}

\listrefs
\end